\documentclass{article}%
\usepackage{amsmath}
\usepackage{amsfonts}
\usepackage{amssymb}
\usepackage{graphicx}%
\providecommand{\U}[1]{\protect\rule{.1in}{.1in}}

\begin{document}

\title{Relational Entropic Dynamics of Particles\thanks{Presented at MaxEnt 2015, the
35th International Workshop on Bayesian Inference and Maximum Entropy Methods
in Science and Engineering (July 19--24, 2015, Potsdam NY, USA). }}
\author{Selman Ipek and Ariel Caticha\\{\small Department of Physics, University at Albany--SUNY, Albany, NY 12222,
USA}}
\date{}
\maketitle

\begin{abstract}
The general framework of entropic dynamics is used to formulate a relational
quantum dynamics. The main new idea is to use tools of information geometry to
develop an entropic measure of the mismatch between successive configurations
of a system. This leads to an entropic version of the classical best matching
technique developed by J. Barbour and collaborators. The procedure is
illustrated in the simple case of a system of $N$ particles with global
translational symmetry. The generalization to other symmetries whether global
(rotational invariance) or local (gauge invariance) is straightforward. The
entropic best matching allows a quantum implementation Mach's principles of
spatial and temporal relationalism and provides the foundation for a method of
handling gauge theories in an informational framework.

\end{abstract}

\section{Introduction}

The question of whether motion is absolute or relative has been around since
the beginning of mechanics. At first the issue was how to define velocity in
order to define motion, but ultimately the issue is how to define change in
general. Newton's solution was to describe motion in terms of the evolving
coordinates of particles embedded in an absolute space and in an absolute time
but objections were immediately raised because, as Newton himself realized,
neither absolute space nor absolute time are observable. A presumably
\textquotedblleft better\textquotedblright\ mechanics would describe motion
not in terms of changes of unobservable absolute positions but in terms of the
observable \emph{relative} distances between them. The search for such
\emph{relational} forms of mechanics eventually led to Einstein's general
relativity and for a while the nature of relative motion was thought to be
fully understood --- at least within the context of classical physics.

Quantum mechanics, however, raises new questions. One problem is that the
quantum version of our best relational theory --- general relativity --- does
not yet exist. Another is that, in its standard formulation, quantum mechanics
is manifestly non-relational: it lives in Newton's absolute space and time, or
at best in Minkowski's absolute space-time. Furthermore, in the standard
Copenhagen interpretation quantum particles do not have definite positions
much less definite relative distances; no such theory could ever be relational
in the usual sense.

Our goal is to take the first steps towards a relational quantum mechanics by
using some of the recent conceptual innovations introduced by Entropic
Dynamics (ED) \cite{Caticha 2010a}-\cite{Caticha 2015a}. ED is a framework for
formulating dynamical models as applications of well established principles of
inference \cite{Caticha 2012} and, as one might expect, the tools for
inference --- probabilities, entropies, and also information geometry --- play
a dominant role. It is indeed a very appealing aspect of ED that its very
foundation emphasizes the close relationship between quantum theory, geometry,
and inference. This suggests that ED might provide the proper setting for the
unification of quantum theory and gravity. Some preliminary steps in this
direction are already being taken in \cite{Nawaz et al 2015} and \cite{Caticha
2015b}.

As with all applications of entropic methods ED requires that we first specify
the subject that is the goal of our inferences --- the microstates. Then we
must identify the relevant information on the basis of which the inference is
to be carried out --- the constraints. Thus, the first step towards a
relational ED is to specify relational microstates and to this end we make
extensive use of the insights into classical versions of relational mechanics
achieved in the pioneering work of Julian Barbour and his collaborators
\cite{Barbour Bertotti 1982}-\cite{Barbour/Mercati}.

The basic problem is that one cannot go very far in formulating a relational
mechanics in terms of the relative interparticle distances --- it is just not
practical. Thus, on one hand one is forced to rely on particle coordinates,
and on the other hand, the coordinate description fails because it is
redundant. Indeed, two configurations that differ by an arbitrary displacement
or an arbitrary rotation describe exactly the same physical situation.

To handle the redundancy Barbour and Bertotti invented a technique they called
best matching (BM) \cite{Barbour Bertotti 1982}. The idea is to introduce some
quantitative measure of the mismatch or the \textquotedblleft
distance\textquotedblright\ between two successive configurations and then
shift and rotate them to find the position of one configuration
\emph{relative} to the other that minimizes the mismatch. Then the actual
intrinsic change is defined through the least mismatch between successive
configurations as they are subjected to translations and rotations relative to
each other.

The Barbour-Bertotti best matching procedure is crucial to implement spatial
relationalism (Mach's first principle) and also temporal relationalism (Mach's
second principle). The former abolishes absolute space. The latter, which
abolishes absolute time, is the notion that physical changes do not occur
\emph{in an external absolute time}, but rather that time is an abstraction at
which we arrive from studying the changes \emph{in things}.

The choice of mismatch measure is central to the relational program. The
particular choice adopted in \cite{Barbour Bertotti 1982} for a relational
classical mechanics is borrowed from Jacobi's action principle and amounts to
a variation on a least-squares mismatch.\ It succeeds and it is elegant but
suffers from the same flaws that one can attribute to other classical
theories: Where do these action principles come from? What is distance? Why
geometry? How does one justify the ad hoc quantization rules that lead to
quantum mechanics? How does one implement best matching in a quantum context?

The ED framework provides insight into all these questions. ED leads us
directly to a quantum theory; no questionable preliminary detours through
classical dynamics are needed. In ED, dynamics is not derived from an
underlying --- and therefore unjustified --- action principle; instead the
action principle is derived. Furthermore, since the quantum states are
probability distributions we already have a unique measure of mismatch. It is
given by the relative entropy and the resulting mismatch is exactly the
distance given by information geometry. This leads us to a criterion of
\emph{entropic best matching} (EBM) which, as we shall see, is all we need to
derive a relational ED.

We conclude this introduction with the observation that the relational program
--- the implementation of Mach's principles through best matching --- can be
applied to more than just the relational motion of particles in space, as Mach
had originally intended. An important insight by Barbour and Bertotti is that
it applies to any model with redundancy in description. Such models include
all fundamental theories such as electromagnetism \cite{Barbour Bertotti
1982}, Yang-Mills theories \cite{Gryb 2009}, and gravity
\cite{Barbour/Mercati}.

In this paper we develop the basic formalism for a relational ED and we
demonstrate entropic best matching in the context of a simple model. We
consider a system of $N$ particles with the simplest redundancy --- a global
translational symmetry. The generalization to include global rotations is
straightforward and will be treated elsewhere.

\section{Entropic Dynamics}

\paragraph{The microstates}

We deal with a system of $N$ particles living in a three dimensional flat
space $\mathbf{X}$ with metric $\delta_{ab}$. We describe the microstate of
the $N$ particle system by a point $x\in\mathbf{X}_{N}$ in a $3N$ dimensional
configuration space, $\mathbf{X}_{N}=\mathbf{X}\otimes\dots\otimes\mathbf{X}$,
with coordinates $x_{n}^{a}$ where $a=1,2,3$ is the spatial index and
$n=1\ldots N$ labels the particle. (Throughout we use notation consistent with
that of \cite{Bartolomeo et al 2014}). The specification of the microstate in
terms of the coordinates $x_{n}^{a}$ is redundant in the sense that shifting
all particles by the same amount does not lead to a different physical state.
In other words, $x_{n}^{a}$ and $x_{n}^{a}+\xi^{a}$ where $\xi^{a}$ is a
constant independent of $x$ and of $n$ describe the same physical situation.

In ED particles are assumed to have well defined positions and the goal is to
infer what those unknown positions might be; we want to assign a probability
distribution $\rho(x)$. The main assumption is that the motion is continuous
which means that it can be analyzed as a sequence of short steps from
$x_{n}^{a}$ to $x_{n}^{a}+\Delta x_{n}^{a}$. The method of maximum entropy is
used to find the probability $P(x^{\prime}|x)$ that the system will take a
short step from $x_{n}^{a}$ to $x_{n}^{\prime a}=x_{n}^{a}+\Delta x_{n}^{a}$.
Then these short steps will be iterated to find the evolving $\rho(x,t)$.

However, in a relational dynamics $x_{n}^{a}$ and $x_{n}^{a}+\xi^{a}$
represent the same initial state just as $x_{n}^{\prime a}$ and $x_{n}^{\prime
a}+\xi^{\prime a}$ represent the same final state. Then the short step is
represented by
\begin{equation}
\hat{\Delta}x_{n}^{a}=(x_{n}^{\prime a}+\xi^{\prime a})-(x_{n}^{a}+\xi
^{a})=\Delta x_{n}^{a}+\Delta\xi^{a}, \label{Delta x}%
\end{equation}
where the spatial shift $\Delta\xi^{a}$ is arbitrary. The two configurations
are said to be \textquotedblleft best matched\textquotedblright\ when
$\Delta\xi^{a}$ is chosen to minimize a certain entropic measure of mismatch
to be defined later. Finding the optimal $\Delta\xi^{a}$ amounts to
establishing a criterion of \textquotedblleft equilocality\textquotedblright%
\ between successive instants; it amounts to deciding which position
$x^{\prime a}$ at the later instant \emph{is that same as}\ a position $x^{a}$
at the earlier instant. Once equilocality has been established two successive
configurations are intrinsically identical when $\hat{\Delta}x_{n}^{a}=0$.

At this point in our argument the optimal $\Delta\xi^{a}$ is still unknown.
However, to make progress we will assume that a notion of equilocality\ has
been established through a trial shift $\Delta\xi^{a}$ to be determined later
when the entropic measure of mismatch is defined.

\paragraph*{Maximum Entropy}

To find the transition probability we maximize the entropy of $P\left(
x^{\prime}|x\right)  $ relative to a prior $Q\left(  x^{\prime}|x\right)  $,%
\begin{equation}
S[P,Q]=-\int dx^{\prime}P\left(  x^{\prime}|x\right)  \log\frac{P\left(
x^{\prime}|x\right)  }{Q\left(  x^{\prime}|x\right)  }\ ,\label{EntPQ}%
\end{equation}
subject to the appropriate constraints. To represent an initial state of
extreme ignorance we adopt a uniform prior.\footnote{Improper,
non-normalizable priors are known to lead to problems. By \textquotedblleft
uniform\textquotedblright\ in a physics context we mean a distribution
$Q(x^{\prime}|x)$ that is sufficiently broad that its precise functional form
does not affect our inferences. In curved spaces a uniform distribution is
such that it assigns equal probabilities to equal volumes. In such cases
$Q(x^{\prime}|x)=(\det g_{AB})^{1/2}f$ $(x^{\prime}|x)$ where $g_{AB}$ is the
metric tensor and $f$ is a sufficiently broad scalar function.} Thus $Q\left(
x^{\prime}|x\right)  \approx Q$ is a constant which can be dropped because it
has no effect on the maximization.

The information about the motion is introduced through constraints. The fact
that particles move by taking infinitesimally short steps is imposed through
$N$ independent constraints,%

\begin{equation}
\left\langle \delta_{ab}\hat{\Delta}x_{n}^{a}\hat{\Delta}x_{n}^{b}%
\right\rangle =\int dx^{\prime}\,P\left(  x^{\prime}|x\right)  \left(
\hat{\Delta}x_{n}^{a}\hat{\Delta}x_{n}^{b}\delta_{ab}\right)  =\kappa
_{n}~.\label{Constraint1}%
\end{equation}
To ensure the continuity of the motion we shall eventually take the limit
$\kappa_{n}\rightarrow0$. Correlations and entanglement among the particles
are imposed through one additional constraint,
\begin{equation}
\sum_{n}\left\langle \hat{\Delta}x_{n}^{a}\right\rangle \frac{\partial
\phi\left(  x\right)  }{\partial x_{n}^{a}}=\int dx^{\prime}P\left(
x^{\prime}|x\right)  \sum_{n}\hat{\Delta}x_{n}^{a}\frac{\partial\phi\left(
x\right)  }{\partial x_{n}^{a}}=\kappa^{\prime},\label{Constraint2}%
\end{equation}
where $\phi$ is a \textquotedblleft drift\textquotedblright%
\ potential\footnote{Elsewhere, in the context of particles with spin, we see
that the potential $\phi(x)$ can be given a natural geometric interpretation
as an angular variable. Its integral over any closed loop is $%
{\displaystyle\oint}
d\phi=2\pi n$ where $n$ is an integer.} which plays a role somewhat analogous
to the pilot wave in de Broglie-Bohm theory and $\kappa^{\prime}$ is another
small but for now unspecified position-independent constant.

Maximizing the entropy (\ref{EntPQ}) subject to the constraints
(\ref{Constraint1}), (\ref{Constraint2}), and normalization leads, after some
manipulation, to a Gaussian process,%

\begin{equation}
P\left(  x^{\prime}|x\right)  =\frac{1}{Z}\exp-\sum_{n}\frac{\alpha_{n}}%
{2}\delta_{ab}\left(  \hat{\Delta}x_{n}^{a}-\frac{\alpha^{\prime}}{\alpha_{n}%
}\frac{\partial\phi}{\partial x_{n}^{a}}\right)  \left(  \hat{\Delta}x_{n}%
^{b}-\frac{\alpha^{\prime}}{\alpha_{n}}\frac{\partial\phi}{\partial x_{n}^{b}%
}\right)  .\label{TransProb}%
\end{equation}
where $Z$ is a normalization constant and $\left\{  \alpha_{n},\alpha^{\prime
}\right\}  $ are Lagrange multipliers. Continuity is achieved imposing that
$\alpha_{n}\rightarrow\infty$. As discussed in \cite{Bartolomeo Caticha 2015}
the multiplier $\alpha^{\prime}$ can be absorbed into $\phi$, which amounts to
setting $\alpha^{\prime}=1$ without changing the dynamics.

\paragraph*{Entropic Time}

In a relational approach time is defined as an ordered succession of instants
but these are not envisaged as being embedded in an externally given absolute
time or space-time. Instead an instant is defined by an \textquotedblleft
instantaneous configuration\textquotedblright. In ED, which has from the very
start been designed to be temporally relational an instant \emph{is defined
by} a probability distribution. In fact, an instant is a probability
distribution \cite{Caticha 2010a}.

The foundation of any notion of time is dynamics and here the dynamics is
given by the transition probability $P\left(  x^{\prime}|x\right)  $ in
(\ref{TransProb}). Thus, if the distribution $\rho(x,t)$ refers to one instant
$t$, then the distribution
\begin{equation}
\rho\left(  x^{\prime},t\right)  =\int dx\,P\left(  x^{\prime}|x\right)
\rho\left(  x,t\right)  \label{DefTime}%
\end{equation}
generated by $P\left(  x^{\prime}|x\right)  $ defines what we mean by the
\textquotedblleft next\textquotedblright\ instant. In ED time is
\emph{constructed} instant by instant so that, given the present, the future
is independent of the past.\footnote{We can see that by construction the
dynamics is Markovian. But this is not the usual Markovian process that occurs
in a pre-existing time; it is an entropic process that generates its own
Markovian time as it unfolds.} The construction leads to instants that are
ordered and, because the transition probability $P\left(  x^{\prime}|x\right)
$ is determined by \emph{maximizing} an entropy, there is a natural arrow of
time. We emphasize that this entropic time is already fully relational: time
is the sequence of instants, time is the sequence of probability distributions.

To complete the construction of entropic time we address what is perhaps the
least fundamental aspect of time: we specify the scale of time $t$. This
amounts to specifying the interval $\Delta t$ between successive instants. The
criterion is convenience: time is defined so as to simplify the description of
motion. For short steps the motion is dominated by fluctuations the scale of
which is given by the multipliers $\alpha_{n}$. This suggests\ setting\
\begin{equation}
\alpha_{n}=\frac{m_{n}}{\hbar\Delta t}~, \label{DefDuration}%
\end{equation}
where $m_{n}$ are particle specific constants that are eventually identified
as masses, and $\hbar$ is an overall constant that fixes the units of time
relative to those of length and mass.

In a relational classical mechanics it is the free particles that provide the
prototype of a clock: particles move equal distances in equal times. In ED it
is the fluctuations that provide the clock that sets the measure of time:
particles undergo equal fluctuations in equal times. With this definition our
Gaussian process eq.(\ref{TransProb}) becomes a Wiener process.

\paragraph*{Short steps}

A generic displacement can be split into the expected drift plus a
fluctuation,
\begin{equation}
\Delta x^{A}=\hat{\Delta}x^{A}-\Delta\xi^{A}=\left\langle \Delta
x^{A}\right\rangle +\Delta w^{A},
\end{equation}
where the capital index $A=(n,a)$ includes both the particle $n$ and the
spatial index $a$ so that $x_{n}^{a}=x^{A}$. To simplify the notation we write
the spatial shift as $\xi^{A}=\xi^{a}$ noting that $\xi^{A}$ is independent of
the particle label $n$. We also introduce the mass tensor,
\begin{equation}
m_{AB}=m_{n}\delta_{AB}\quad\text{and its inverse}\quad m^{AB}=\frac{1}{m_{n}%
}\delta_{AB}~,
\end{equation}
which, as shown \cite{Bartolomeo et al 2014}\cite{Caticha 2015a}, is the
metric of configuration space up to an unimportant scale factor.

From (\ref{Delta x}) and (\ref{TransProb}) we find that the shift $\Delta
\xi^{A}$ affects the expected steps,
\begin{equation}
\left\langle \Delta x^{A}\right\rangle =\left\langle \hat{\Delta}%
x^{A}\right\rangle -\langle\Delta\xi^{A}\rangle=\hbar\Delta tm^{AB}%
\partial_{B}\phi-\Delta\xi^{A}\ ,\label{ExpX}%
\end{equation}
but does not affect the fluctuations,
\begin{equation}
\hat{\Delta}w^{A}=\Delta w^{A}\quad\text{with}\quad\left\langle \Delta
w^{A}\right\rangle =0\quad\text{and\quad}\left\langle \Delta w^{A}\Delta
w^{B}\right\rangle =\hbar\Delta t\,m^{AB},\label{ExpW}%
\end{equation}
which remain large $\Delta w\sim O(\Delta t^{1/2})$ and essentially
isotropic.\ This leads us to expect, and we shall later confirm, that the
optimal shift $\Delta\xi$ is of order $\Delta t$ so that $\Delta\xi\ll\Delta
w$.

\paragraph{Fokker-Planck Equation}

The dynamical equation of evolution, eq.(\ref{DefTime}) can be rewritten as a
Fokker-Planck (FP) or continuity equation \cite{Caticha 2012},
\begin{equation}
\partial_{t}\rho\left(  x,t\right)  =-\partial_{A}\left[  \rho\left(
x,t\right)  V^{A}\left(  x,t\right)  \right]  ~,\label{FP a}%
\end{equation}
where $V^{A}$ is the velocity of the probability flow, or current velocity,%
\begin{equation}
V^{A}\left(  x,t\right)  =m^{AB}\partial_{B}\Phi\left(  x,t\right)  -\dot{\xi
}^{A}\text{\quad where\quad}\Phi=\hbar(\phi-\log\rho^{1/2}%
)~,\label{CurrentVel}%
\end{equation}
and $\dot{\xi}^{A}=\Delta\xi^{A}/\Delta t$.

The current velocity receives three types of contributions. The first two are
the familiar drift and osmotic velocities described through the gradient of
the \textquotedblleft phase\textquotedblright\ $\Phi$ \cite{Caticha 2015a}.
The third contribution is the shift velocity $\dot{\xi}^{A}$; it is the term
that implements relationality.

\section{Entropic Best Matching}

We are now ready to introduce entropic best matching (EBM). The basic idea is
that once we have decided on the relevant information necessary for predicting
future behavior we can say that this information defines what we mean by an
\textquotedblleft instant\textquotedblright. In ED the relevant information is
given by the distribution $\rho(x,t)$ and the (suitably updated) drift
potential $\phi(x,t)$ which determines the transition probability
$P(x^{\prime},t^{\prime}|x,t)$. An alternative representation of exactly the
same information is given by the \emph{joint} distribution\footnote{This is
not the probability of $t$ and $t^{\prime}$; perhaps a better notation would
be $\rho(x^{\prime},x|t^{\prime},t)$ where $t$ and $t^{\prime}$ are
parameters.}
\begin{equation}
\rho(x^{\prime},t^{\prime};x,t)=P(x^{\prime},t^{\prime}|x,t)\rho(x,t)~.
\label{joint}%
\end{equation}
This representation is more convenient for two reasons. First, the joint
distribution (\ref{joint}) is more suggestive of the flow of time from $t$ to
$t^{\prime}$. Indeed, while the pair of functions $\rho(x,t)$ and $\phi(x,t)$
are assigned to a single sharp instant $t$, the single function $\rho
(x^{\prime},t^{\prime};x,t)$ refers to two instants $t$ and $t^{\prime}$. In
fact, $\rho(x^{\prime},t^{\prime};x,t)$ describes evolution from the initial
instant $t$ not just into one other instant $t^{\prime}$ but into all instants
in the immediate future of $t$. This is precisely what we need to explore the
dynamical effect of any trial shift $\Delta\xi^{A}$.

Second, and equally significant, is the fact that $\rho(x^{\prime},t^{\prime
};x,t)$ is a probability distribution while $\phi(x,t)$ is not. This means
that there exists a unique criterion for quantifying the mismatch between any
two such distributions; it is given by their relative entropy. For
distributions that differ only slightly, as we expect in the case of
\emph{successive} instants, this is equivalent to the information distance
between them.\footnote{For reviews on information geometry including the proof
of uniqueness of the information metric, see \emph{e.g.}, \cite{Caticha
2012}\cite{Amari 1985}.} Therefore the mismatch that is relevant to our
discussion is the information distance between $\rho\left(  x^{\prime
},t^{\prime};x,t\right)  $ and $\rho\left(  x^{\prime},t^{\prime
}+dt;x,t\right)  $.

To summarize: A natural implementation of EBM within a framework of inference
requires that we first identify which distributions are to compared; then,
using their relative entropy, we quantify their mismatch as an information
distance. The \emph{intrinsic} mismatch is the minimum distance between the
distributions as they are shifted relative to each other by varying the shift
$\Delta\xi^{A}$. The optimal shift $\Delta\xi^{A}$ implements equilocality.

\paragraph*{Information Geometry and Temporal Distance}

The information distance $dT$ between $\rho\left(  x^{\prime},t^{\prime
};x,t\right)  $ and $\rho\left(  x^{\prime},t^{\prime}+dt;x,t\right)  $ is
given by $dT^{2}=Gdt^{2}$ where the metric tensor $G$ has a single component,
\begin{equation}
G=C\int dxdx^{\prime}\rho\left(  x^{\prime},t^{\prime}|x,t\right)  \left[
\partial_{t^{\prime}}\log\rho\left(  x^{\prime},t^{\prime}|x,t\right)
\right]  ^{2}~,
\end{equation}
where $C$ is an arbitrary overall constant. Using (\ref{joint}),
\begin{equation}
G=C\int dx\rho\left(  x,t\right)  \int dx^{\prime}P\left(  x^{\prime
},t^{\prime}|x,t\right)  \left[  \partial_{t^{\prime}}\log P\left(  x^{\prime
},t^{\prime}|x,t\right)  \right]  ^{2},\label{TempMetric}%
\end{equation}
and then (\ref{TransProb}) gives
\begin{equation}
G=\frac{3NC}{2(\Delta t)^{2}}+\frac{C}{\hbar\Delta t}\int dx\rho\left(
x,t\right)  m^{AB}(\hbar\partial_{A}\phi-\dot{\xi}_{A})(\hbar\partial_{B}%
\phi-\dot{\xi}_{B})\,~,
\end{equation}
where $\Delta t=t^{\prime}-t$. The divergence as $\Delta t\rightarrow0$ is a
consequence of the fact that as $\Delta t\rightarrow0$ the distributions
$P\left(  x^{\prime},t^{\prime}|x,t\right)  $ become infinitely narrow and
therefore infinitely distinguishable. The problem is mildly annoying but not
fatal because our interest is not in $G$ itself but in how it changes as we
vary $\Delta\xi^{A}$; it can be alleviated by choosing the arbitrary constant
$C=\hbar\Delta t/2$ so that
\begin{equation}
G=\frac{3N\hbar}{4\Delta t}+\frac{1}{2}\int dx\rho\,m^{AB}(\hbar\partial
_{A}\phi-\dot{\xi}_{A})(\hbar\partial_{B}\phi-\dot{\xi}_{B})~.
\end{equation}
Using (\ref{CurrentVel}) to write $\phi$ in terms of $\Phi$ we find,
\begin{equation}
G=\frac{3N\hbar}{4\Delta t}-\frac{\hbar}{2}\dot{S}+\tilde{H}_{0}\left[
\rho,\Phi\right]  ~.\label{TempMetricBM}%
\end{equation}
The first term is not interesting; it is a constant independent of $\dot{\xi
}_{A}$. The second term is a bit more interesting because $\dot{S}$ turns out
to be the rate of entropy increase,%
\begin{equation}
\dot{S}=\frac{dS}{dt}\quad\text{with}\quad S[\rho]=-\int dx\,\rho\left(
x,t\right)  \log\rho\left(  x,t\right)  ~.
\end{equation}
Its contribution to the metric $G$ is what one might expect: when $\dot{S}>0$
the distributions are broader, more difficult to discriminate, and the
information distance between them decreases. However $\dot{S}$ is also
independent of $\dot{\xi}_{A}$ and does not contribute to best matching. The
explicit expression is
\begin{equation}
\dot{S}=-\int dx\rho\,\,\partial^{A}\left(  \partial_{A}\Phi-\dot{\xi}%
_{A}\right)  =-\int dx\rho\,\,\partial^{A}\partial_{A}\Phi~,
\end{equation}
where we used $\partial_{A}\dot{\xi}_{B}\left(  t\right)  =0$. The term in
(\ref{TempMetricBM}) that is crucial for EBM is the last one,
\begin{equation}
\tilde{H}_{0}\left[  \rho,\Phi\right]  =\int dx\left[  \frac{m^{AB}}{2}%
\rho\left(  \partial_{A}\Phi-\dot{\xi}_{A}\right)  \left(  \partial_{B}%
\Phi-\dot{\xi}_{B}\right)  +\frac{\hbar^{2}}{8}\frac{1}{\rho}\partial_{A}%
\rho\partial_{B}\rho\right]  ~,\label{H_0}%
\end{equation}
which can be recognized as the ensemble Hamiltonian (minus a potential energy
term). Thus, considerations of information geometry have lead to a metric that
includes the kinetic energy and the quantum potential. The significance of
this finding will be more fully explored elsewhere.

\paragraph*{Entropic Best Matching}

Entropic BM is achieved by minimizing the mismatch between $\rho\left(
x^{\prime},t^{\prime};x,t\right)  $ and $\rho\left(  x^{\prime},t^{\prime
}+dt;x,t\right)  $ as measured by the distance $dT$. This amounts to
minimizing $G$ with respect to trial shifts $\dot{\xi}^{a}$. The result of
this variation is the constraint:%

\begin{equation}
\frac{\partial G}{\partial\dot{\xi}^{a}}=\int dx\rho%
{\displaystyle\sum\nolimits_{n}}
\frac{\partial\Phi}{\partial x_{n}^{a}}-M\,\dot{\xi}_{a}^{\text{best}}=0,
\label{BMConstraint}%
\end{equation}
where we have used that $\dot{\xi}^{A}=\dot{\xi}^{a}$ is independent of $x$
and of $n$ and $M=\sum_{n}m_{n}$. Incidentally, (\ref{BMConstraint}) confirms
our earlier intuition that $\Delta\xi^{\text{best}}\sim O(\Delta t)$.

The interpretation of $\dot{\xi}_{a}^{\text{best}}$ is straightforward. As
discussed in \cite{Caticha 2015a} the integral
\begin{equation}
\tilde{P}_{a}=\int d^{3N}x\,\rho%
{\displaystyle\sum\nolimits_{n}}
\frac{\partial\Phi}{\partial x_{n}^{a}}=\int d^{3N}x\,\rho\frac{\partial\Phi
}{\partial X^{a}}%
\end{equation}
is interpreted as the expectation of the total momentum, and $X^{a}$ are the
coordinates of the center of mass,
\begin{equation}
X^{a}=\frac{1}{M}%
{\displaystyle\sum\nolimits_{n}}
m_{n}x_{n}^{a}~.
\end{equation}
Equation (\ref{BMConstraint}) can now be read in two ways. If we are given a
sequence of consecutive states $\{\rho,\Phi\}$ the shift $\dot{\xi}%
_{a}^{\text{best}}$ that achieves equilocality is given by the velocity of the
center of mass,
\begin{equation}
M\dot{\xi}_{a}^{\text{best}}=\tilde{P}_{a}~.
\end{equation}
Alternatively, we can require that consecutive states be best matched, that
is, no shift is needed: $\dot{\xi}_{a}^{\text{best}}=0$. This means that
successive states $\{\rho,\Phi\}$ are constrained to satisfy $\tilde{P}_{a}%
=0$. In this second reading (\ref{BMConstraint}) is a constraint on the
allowed states.

\section{Relational Quantum Dynamics}

The dynamics described by eq.(\ref{FP a}) for any externally prescribed drift
potential $\phi$ is a diffusion; it is an unusual diffusion in that it is
relational, but it is not yet a \emph{quantum} theory. The way to a quantum
mechanics is to require that the diffusion be non-dissipative which is what
leads to a Hamiltonian dynamics.

The idea is to recognize that, just as $\rho$ evolves in response to $\phi$,
we must also allow $\rho$ to react back so that $\phi$ evolves in response to
$\rho$. This promotes $\varphi$, or equivalently $\Phi$, to a dynamical
variable. The precise recipe for this coupled evolution is to require that
there be some conserved quantity $\tilde{H}\left[  \rho,\Phi\right]  $ such
that changes in $\rho$ induce changes in $\Phi$ in a way that $\tilde
{H}\left[  \rho,\Phi\right]  $ remains constant. The quantity $\tilde
{H}\left[  \rho,\Phi\right]  $ is chosen so that the evolution of $\rho$ is
described by eq.(\ref{FP a}). The result is the coupled evolution of $\rho$
and $\Phi$ given by Hamilton's equations \cite{Bartolomeo et al 2014},%

\begin{equation}
\partial_{t}\rho=\frac{\delta\tilde{H}\left[  \rho,\Phi\right]  }{\delta\Phi
}\quad\text{and\quad}\partial_{t}\Phi=-\frac{\delta\tilde{H}\left[  \rho
,\Phi\right]  }{\delta\rho}. \label{HamEqns}%
\end{equation}
The Hamiltonian $\tilde{H}\left[  \rho,\Phi\right]  $ that reproduces
(\ref{FP a}) is of the form,%
\begin{equation}
\tilde{H}\left[  \rho,\Phi\right]  =\int dx\,\frac{\rho}{2}m^{AB}\left(
\partial_{A}\Phi-\dot{\xi}_{A}\right)  \left(  \partial_{B}\Phi-\dot{\xi}%
_{B}\right)  +F\left[  \rho\right]  , \label{GenHam}%
\end{equation}
where the functional $F\left[  \rho\right]  $ is an integration constant. The
second Hamilton equation is recognized as a Hamilton--Jacobi (HJ) equation,
\begin{align}
-\partial_{t}\Phi &  =\frac{1}{2}m^{AB}\left(  \partial_{A}\Phi-\dot{\xi}%
_{A}\right)  \left(  \partial_{B}\Phi-\dot{\xi}_{B}\right)  +\frac{\delta
F}{\delta\rho}\nonumber\\
&  =\sum_{n}\frac{\delta^{ab}}{2m_{n}}\left(  \frac{\partial\Phi}{\partial
x_{n}^{a}}-m_{n}\dot{\xi}_{a}\right)  \left(  \frac{\partial\Phi}{\partial
x_{n}^{b}}-m_{n}\dot{\xi}_{b}\right)  +\frac{\delta F}{\delta\rho}~.
\end{align}
$\partial\Phi/\partial x_{n}^{a}$ is interpreted as the momentum $p_{a}$ of
the $n^{\text{th}}$ particle. In fact, $p_{a}$ is the \textquotedblleft
coordinate\textquotedblright\ momentum; the intrinsic momentum that reflects
true change is the momentum corrected by the shift velocity, $p_{a}-m_{n}%
\dot{\xi}_{a}$.\footnote{Something analogous occurs in electromagnetism. The
canonical momentum $p^{a}$ is split into a kinetic momentum $\pi^{a}$ and the
momentum contained in the electromagnetic field, $eA^{a}/c$. Therefore
$p^{a}=\pi^{a}+eA^{a}/c$. The momentum $mv^{a}$ that contributes to the
kinetic energy $mv^{2}/2$ is $mv^{a}=\pi^{a}=p^{a}-eA^{a}/c$.}

The advantage of EBM over classical BM is that the framework introduced here
is quantized in a straightforward fashion. In ED quantum theory results from a
particular choice of the functional $F\left[  \rho\right]  $. What is
especially interesting is that this special choice is suggested by the
information metric in eq.(\ref{TempMetricBM}) and specifically by the term
$\tilde{H}_{0}$, eq.(\ref{H_0}). By choosing $F\left[  \rho\right]  $ so that
the Hamiltonian coincides with eq.(\ref{H_0}) we indeed obtain a quantum
theory, a theory of $N$ free particles. More interesting quantum theories are
obtained by the inclusion of some scalar potential $U\left(  x\right)  $. This
results in the quantum Hamiltonian,
\begin{equation}
\tilde{H}\left[  \rho,\Phi\right]  =\tilde{H}_{0}\left[  \rho,\Phi\right]
+\int dx\,\rho U~. \label{QuantHam1}%
\end{equation}

The fact that this is now a \emph{quantum} theory can be made explicit by
combining $\rho$ and $\Phi$ into a complex wave function $\Psi=\rho^{1/2}%
\exp\left(  i\Phi/\hbar\right)  $. Then Hamilton's equations (\ref{HamEqns})
can be combined into a single linear Schr\"{o}dinger equation,
\begin{equation}
i\hbar\partial_{t}\Psi=\frac{1}{2}m^{AB}\left(  i\hbar\partial_{A}-\dot{\xi
}_{A}\right)  \left(  i\hbar\partial_{B}-\dot{\xi}_{B}\right)  \Psi+U\left(
x\right)  \Psi. \label{SchroEqn}%
\end{equation}

If we require that consecutive states be best matched, that is we impose
$\dot{\xi}_{a}=0$, then the states $\{\rho,\Phi\}$ are constrained to satisfy
$\tilde{P}_{a}=0$. In terms of $\Psi$ this constraint takes the form
\begin{equation}
\int dx\,\Psi^{\ast}\left(  x,t\right)
{\displaystyle\sum\nolimits_{n}}
\hat{p}_{na}\Psi\left(  x,t\right)  =\langle\hat{P}_{a}\rangle=0,
\label{QBMConstraint}%
\end{equation}
where $\hat{p}_{na}=i\hbar\partial/\partial x_{n}^{a}$ is the quantum momentum
operator for particle $n$ and $\hat{P}_{a}=\sum_{n}\hat{p}_{na}$. Thus, EBM
imposes a\textit{\ }constraint that restricts solutions to the subspace of the
full Hilbert space where the expected total momentum vanishes.

As a side remark, it is interesting that EBM imposes \textit{weak}
constraints, or rather, expected value constraints. In the standard approach
to quantizing theories with constraints, questions arise as to whether
constraints should be imposed on operators, on states, or on expectation
values. EBM provides an answer --- quantum constraints are to be imposed on
expectation values.

To complete the relational ED we require yet another condition, a consistency
condition. For the dynamics to remain relational we must require that the
potential $U\left(  x\right)  $ be such that the constraint $\langle\hat
{P}_{a}\rangle=0$ be preserved by the dynamical evolution. This implies that
$\langle\hat{P}_{a}\rangle$ must be a constant of the motion. This condition
is satisfied if the potential $U\left(  x\right)  $ depends only on the
relative particle positions, $U\left(  x\right)  =U\left(  \{\vec{x}_{i}%
-\vec{x}_{j}\}\right)  $. The non-conservation of momentum would indicate the
existence of a preferred absolute frame of reference.

\section{Discussion and Conclusion}

We have formulated a relational quantum dynamics by implementing an entropic
form of best matching. The framework has been developed for the simple case of
global translations in particle dynamics but it can be easily generalized to
global rotations in which case the imposed constraint is that the total
expected angular momentum vanish. That these results are statistical analogues
of the classical BM is both encouraging and deceiving. On one hand, one
expects EBM to reduce to classical BM in some limit, but the two frameworks
are very different animals. Classical BM compares ontic particle
configurations while EBM compares epistemic probability distributions. The
difference is highlighted in the case of a single particle. Classically we
cannot even ask about the relational motion of a single particle. Relative to
what would that motion happen? The situation in EBM is appreciably different,
there is no problem either in principle or in practice for even a single
particle because it is probability distributions that are being best-matched
and not the particle configurations. Another important difference is that
classical BM relies on the classical action as a best-matching criterion, EBM
relies on a purely inferential criterion provided by information geometry. In
fact, while classical BM\ requires previous knowledge of the dynamics, in this
work we have uncovered hints that EBM will help us find the Hamiltonian that
induces the dynamics.

To conclude, the theory developed here is not yet fully relational. This is
only a first step towards a relational framework that applies to the local
gauge symmetries including diffeomorphisms and local dilatations that are
central to all fundamental theories such as electromagnetism, Yang-Mills
theories, and gravity.

\subsubsection*{Acknowledgements}

We would like to thank M. Abedi, D. Bartolomeo, C. Cafaro, N. Caticha, S.
DiFranzo, A. Giffin, D.T. Johnson, K. Knuth, S. Nawaz, M. Reginatto, C.
Rodr\'{\i}guez, and K. Vanslette for many discussions on entropy, inference
and quantum mechanics.

\end{document}